\documentclass[conference,10pt]{IEEEtran}
\usepackage{graphicx,subfigure}
\usepackage{amsmath,graphicx}
\usepackage{amsmath,amssymb,amsfonts}
\usepackage{cite}
\usepackage{url}
\usepackage{bm}
\usepackage{verbatim}
\usepackage{float}
\usepackage{subfig}

\usepackage[usenames,dvipsnames]{color}

\title{Space-Time Design for Deep Joint Source Channel Coding of Images over MIMO Channels}
 \author{%
   \IEEEauthorblockN{Chenghong Bian, Yulin Shao, Haotian Wu, and Deniz G{\"u}nd{\"u}z}
   
   \IEEEauthorblockA{Department of Electrical and Electronic Engineering, Imperial College London, London SW7 2BT, UK
   \\
    Email:$\{$c.bian22, y.shao, haotian.wu17, d.gunduz$\}$@imperial.ac.uk
    }
   
}

\begin{document}
%

\newcommand{\figstructure}{
  \begin{figure}
    \centering
    \includegraphics[width=\linewidth]{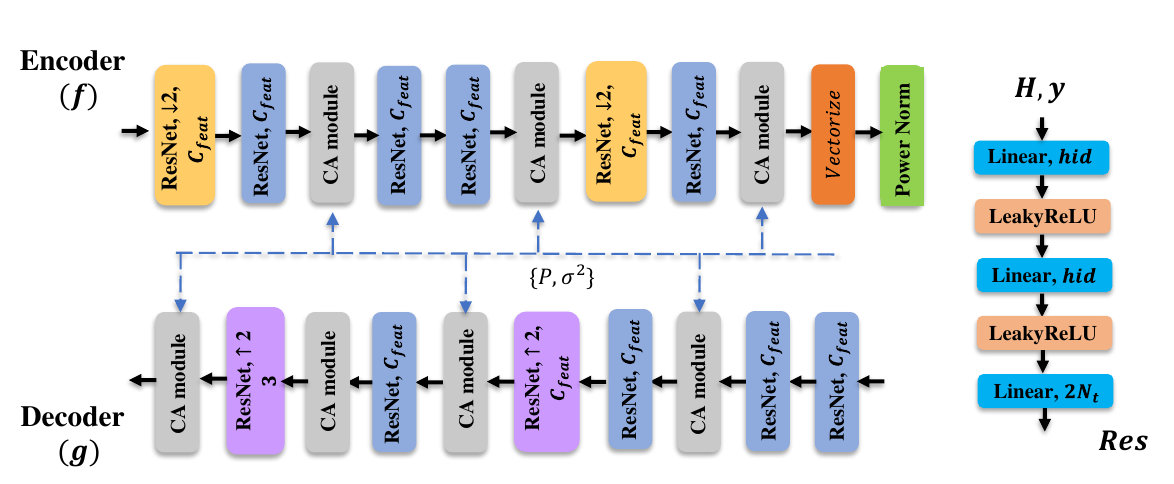}
    \caption{The structure of the encoder (top), decoder (bottom), and the residual block for MIMO equalization at the receiver (right). The CA modules take features, signal power, $P$, and noise power $\sigma^2$ as input. $hid$ denotes the number of hidden neurons in the residual block.}
    \label{fig:fig_structure}
  \end{figure}
}

\newcommand{\figtradeoff}{
  \begin{figure}
    \centering
    \includegraphics[width=0.8\linewidth]{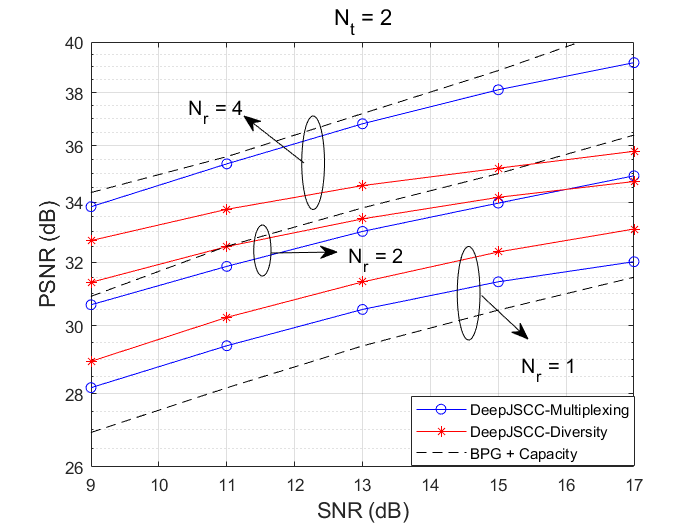}
    \caption{The PSNR performance of the proposed schemes and the digital baseline with $N_t=2, N_r \in \{1,2,4\}$.}
    \label{fig:fig_tradeoff}
  \end{figure}
}

\newcommand{\figNt}{
  \begin{figure}
    \centering
    \includegraphics[width=0.8\linewidth]{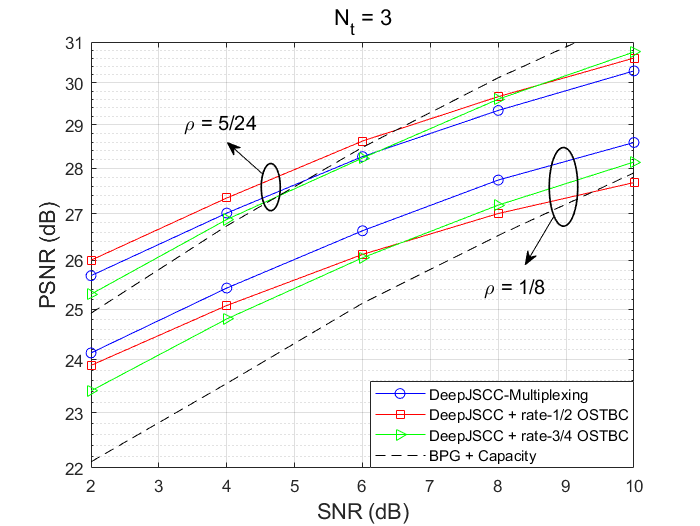}
    \caption{The PSNR performance with $N_t=3, N_r =1$, and \textit{bandwidth ratios}, $\rho = 1/8, 5/24$,.}
    \label{fig:fig_nt3}
  \end{figure}
}

\newcommand{\figSINR}{
  \begin{figure}
    \centering
    \includegraphics[width=0.8\linewidth]{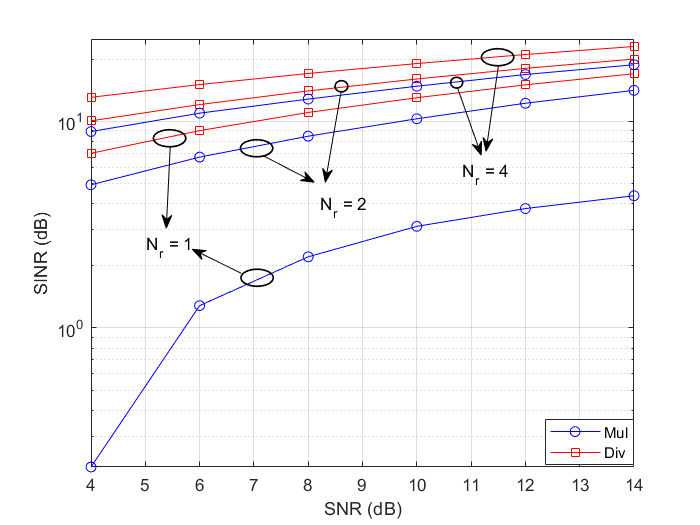}
    \caption{SINR comparison among the proposed schemes for different number of receive antennas.}
    \label{fig:fig_SINR}
  \end{figure}
}

\newcommand{\figsystem}{
\begin{figure}
     \centering
     \begin{subfigure}
         \centering
         \includegraphics[width=\columnwidth]{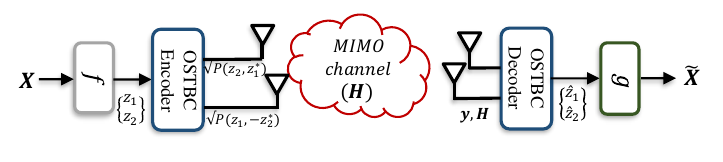}
         \centerline{(a) Full diversity transmission.}\medskip
     \end{subfigure}
     \begin{subfigure}
         \centering
         \includegraphics[width=\columnwidth]{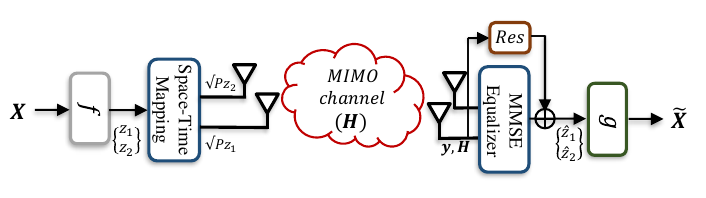}
         \centerline{(b) Multiplexing scheme.}\medskip
     \end{subfigure}

  \caption{Illustration of the (a) diversity and (b) multiplexing schemes over a $2 \times 2$ MIMO channel. The former employs the Alamouti scheme to achieve full diversity. }
\label{fig:fig_system}
\end{figure}}

\maketitle

\begin{abstract}
We propose novel deep joint source-channel coding (DeepJSCC) algorithms for wireless image transmission over multi-input multi-output (MIMO) Rayleigh fading channels, when channel state information (CSI) is available only at the receiver. We consider two different schemes; one exploiting the spatial diversity and the other exploiting the spatial multiplexing gain of the MIMO channel, respectively. For the former, we utilize an orthogonal space-time block code (OSTBC) to achieve full diversity and increase the robustness against channel variations. In the latter, we directly map the input to the antennas, where the additional degree of freedom can be used to send more information about the source signal. Simulation results show that the diversity scheme outperforms the multiplexing scheme for lower signal-to-noise ratio (SNR) values and a smaller number of receive antennas at the AP. When the number of transmit antennas is greater than two, however, the full-diversity scheme becomes less beneficial. We also show that both the diversity and multiplexing schemes can achieve comparable performance with the state-of-the-art BPG algorithm delivered at the instantaneous capacity of the MIMO channel, which serves as an upper bound on the performance of separation-based practical systems.
\end{abstract}
\section{Introduction}
\label{sec:intro}

Shannon's separation theorem states that separate source and channel coding is optimal at the infinite block length regime. However, practical systems are limited to finite block lengths, and separation can be highly sub-optimal. Over the years, researchers have designed many joint source and channel coding (JSCC) algorithms to improve the reconstruction performance at the receiver \cite{trad_jscc}. However, due to the difficulty of designing and implementing such codes, separation-based schemes are still widely used. Recently, due to the increasing interest in latency-constrained edge applications, JSCC schemes are receiving growing interest, also under the framework of semantic communications \cite{sematic_deniz}.

Deep neural networks (DNNs) have found applications in various communication problems, including compression \cite{image_compress, neuraldsc} and channel coding \cite{sparc,feedbackcode}. The lack of high-performing practical JSCC schemes also motivated research on DNN-aided JSCC schemes \cite{bourtsoulatze2019deep, NTSCC, Kurka:TWC:21, jsccofdm,wu2022vision}. In \cite{bourtsoulatze2019deep}, the authors proposed the first DNN-based JSCC scheme, called DeepJSCC, for wireless image transmission, and showed that it achieves comparable performance with state-of-the-art separation schemes while avoiding the `cliff-effect'. DeepJSCC is later extended to channels with feedback \cite{kurka2020deepjscc}, to successive refinement \cite{Kurka:TWC:21}, and to multi-path fading channels in \cite{jsccofdm, wu2022channel}. All of these works illustrate that DeepJSCC is a promising technology for next-generation communication systems.

JSCC over block-fading multiple-input multiple-output (MIMO) channels is studied in \cite{jsccmimogaussian} from an information-theoretic perspective, focusing on the high signal-to-noise ratio (SNR) regime. The authors proposed a layered transmission scheme, and showed that each layer should operate at a different point on the diversity-multiplexing trade-off \cite{div_mul}. This analysis is later extended to the finite SNR regime in \cite{layered_broadcast}. However, the analysis in both of these papers is mainly theoretical focusing on Gaussian sources.

In this paper, our goal is to design JSCC schemes for practical image transmission over MIMO fading channels when perfect channel state information (CSI) is available only at the receiver.  The most relevant work is \cite{wu2022channel} where the CSI is also assumed to be available at the transmitter which allows the transmitter to perform beamforming according to the CSI and differs from our setting.  In particular, we consider a MIMO transmitter-receiver pair, and propose two schemes. In the first scheme, called the \textit{diversity scheme}, we employ an orthogonal space-time block code (OSTBC) to benefit from the increased diversity provided by the multiple transmit antennas. The encoder first applies a non-linear transform on the input image through a DNN architecture to extract the most important features for the recovery of the input image. Then, the obtained features are mapped to the antennas and the time slots according to the OSTBC design for transmission. The receiver can decouple the received signals at its antennas and achieve full diversity. This introduces redundancy but increases robustness against channel variations. In the second, so-called \textit{multiplexing scheme}, the encoder directly maps the latent vector to the antennas. Using the CSI, the receiver first equalizes the received signal using residual-assisted \cite{jsccofdm, mimo_nos} minimum mean square error (MMSE) equalizer, and then reconstructs the image directly. Note that the latter scheme is more general, and in principle, the DeepJSCC encoder/decoder pair, trained in an end-to-end fashion, can learn to introduce redundancy against channel variations when necessary (e.g., in the low SNR regime). Our goal here is to investigate whether introducing our domain knowledge into the DeepJSCC architecture through the employment of an OSTBC can help the system to achieve better end-to-end performance. Simulation results confirm the benefits of this approach in the low SNR regime, which highlights the importance of domain knowledge and model-based design when incorporating DNNs into the design of future communication systems. Simulation results also reveal the comparable performance of the proposed schemes to the separation scheme using BPG algorithm with instantaneous MIMO capacity, which results in a rather loose upper bound on the performance of any separation-based scheme employing BPG for compression.

\figsystem

\section{Problem Formulation}
We consider the transmission of images over a Rayleigh fading MIMO channel with
$N_t$ transmit and $N_r$ receive antennas. We denote the input image by $\mathbf{X} \in \mathbb{R}^{C\times H \times W}$, where $C$, $H$ and $W$ are the number of color channels, height, and width of the image. The transmitter transforms $\mathbf{X}$ into a matrix $\bm{S} \in \mathbb{C}^{N_t\times k}$ with ${||\bm{S}||^2_F \le N_t k}$, where $k$ denotes the number of channel uses.  We define the \textit{bandwidth ratio}, $\rho$, as the number of channel uses per pixel, i.e., $\rho \triangleq k/(CHW)$. Finally, given the power constraint $P$ at each transmit antenna, $\bm{S}$ is scaled to $\sqrt{P}\bm{S}$ before transmission.

We assume a block Rayleigh fading channel, which remains constant during the transmission of one image, and changes to an independent state for the next. The channel state is denoted by $\bm{H} \in \mathbb{C}^{N_r \times N_t}$, where each independent element follows a complex Gaussian distribution with zero-mean and unit-variance. The received signal can be written as
\begin{equation}
     \bm{Y} = \sqrt{P}\bm{H}\bm{S} + \bm{N},
    \label{eq:channel_model}
\end{equation}
where $\bm{Y} \in \mathbb{C}^{N_r \times k}$, and $\bm{N} \in \mathbb{C}^{N_r \times k}$ denotes the complex additive white Gaussian noise (AWGN), $\bm{N}_{ij} \sim \mathcal{CN}(0,\sigma^2)$. We define the average SNR as:
\begin{equation}
     SNR  \triangleq \frac{P}{N_t\sigma^2},
    \label{eq:SNR}
\end{equation}
where $N_t$ in the denominator ensures that the average SNR at each receive antenna remains constant with different numbers of transmit antennas.
We assume that the receiver has perfect CSI while the transmitter does not have any CSI. 
The goal is to minimize the loss between the original image and the reconstructed one at the receiver, which will be introduced in the subsequent section.

\figstructure
\section{DeepJSCC over a MIMO Channel}
\label{sec:method}
For the proposed DeepJSCC methods, the user first encodes $\mathbf{X}$ by a DeepJSCC encoder, denoted by $f$, and obtain the latent vector $\bm{z} \in \mathbb{C}^l$, which is normalized as:
\begin{equation}
     \mathbb{E}||\bm{z}||_2^2 \leq l,
     \label{eq:pwr_constraint}
\end{equation}
where $l$ denotes the number of complex symbols employed to encode the image before mapping to the antennas.

Before transmission, the transmitter applies a space-time mapping (STM) to the normalized vector, which can be either a space-time coding scheme in order to achieve diversity gain, or an identical mapping to achieve multiplexing gain.
We denote by $\bm{s}$ the output of the STM, which is then loaded to $N_t$ antennas sequentially and scaled to form the transmitted signal $\sqrt{P}\bm{S}$.  After passing the channel as in \eqref{eq:channel_model}, an estimate of the latent vector, denoted by $\bm{\widetilde{z}}$, is generated based on the received signal $\bm{Y}$ and the CSI, $\bm{H}$. Then the reconstructed image can be expressed as $\bm{\widetilde{X}} = g(\bm{\widetilde{z}})$, where $g$ stands for the neural decoder function. The mean squared error (MSE) loss is adopted to measure the reconstruction error, expressed as: $\mathcal{L}(f, g) = \mathbb{E} ||\bm{X} - \bm{\widetilde{X}}||_F^2$.

The encoder ($f$) and decoder ($g$) structures are shown in Fig.~\ref{fig:fig_structure}, where the pixel shuffling \cite{pixel-shuffle} is used inside the module `$ResNet, \uparrow 2$' to upsample input features. We also employ channel attention (CA) modules \cite{attention_jscc, wu2022channel}, which takes $P$ and $\sigma^2$ as additional inputs in both $f$ and $g$ to ensure that the system is robust to different SNR regimes. 

It is well-known that multiple antennas can provide both diversity and multiplexing gain \cite{div_mul}. In separation-based approaches, different space-time codes \cite{ostbc,ldcode} can be exploited to operate on different points of the trade-off between the diversity and multiplexing gains in a MIMO channel. Here, our goal is to explore whether such schemes can also benefit the DeepJSCC approach by explicitly introducing structures into the code design. We will introduce two distinct transmission strategies to achieve this, and explore whether an explicit diversity scheme can be beneficial to achieve a more robust DeepJSCC performance, especially in the low SNR regime.

\subsection{Full Diversity Transmission}
The first scheme aims to achieve full diversity via OSTBC \cite{ostbc}. We will first overview the encoding and decoding process of the OSTBC, and then provide the detailed operations for the $N_t = 2, 3$ cases.

We consider OSTBCs, which fall into the class of \textit{generalized complex orthogonal designs} as a complex channel is assumed. Following \cite{ostbc}, an OSTBC design is denoted by an STM matrix $\mathcal{G}\in \mathcal{C}^{N_t \times N_u}$, where $N_u$ represents the number of channel uses. The number of distinct elements in $\mathcal{G}$ is denoted by $m$, and the code rate is given by $R = m/N_u$.  To match the encoder output, $\bm{z}$, with the OSTBC design, we first partition $\bm{z}$ into multiple tuples $\{\bm{z}^{(1)}, \cdots, \bm{z}^{(k/N_u)}\}$, each of which contains $m$ elements\footnote{Note that $\bm{z}$ can be zero-padded if its length is not a multiple of $m$.}. Then, we map the elements of each tuple to different transmit antennas and time slots according to $\mathcal{G}$, to be transmitted over the MIMO channel. The entire transmitted signal $\bm{S}$ is obtained by stacking $\mathcal{G}(\bm{z}^{(i)})$ for each and every tuple in the time domain:
\begin{equation}
    \bm{S} = \left[\mathcal{G}(\bm{z}^{(1)}), \cdots, \mathcal{G}(\bm{z}^{(k/N_u)})\right].
\end{equation}
The $i$-th received OSTBC block $\bm{Y}^{(i)}\in \mathcal{C}^{N_r \times N_u}$ can be expressed as:
\begin{equation}
    \bm{Y}^{(i)} = \sqrt{P} \bm{H} \mathcal{G}(\bm{z}^{(i)})+ \bm{N},
    \label{eq:received_y}
\end{equation}
which is first processed using the OSTBC decoder at the receiver. As illustrated in \cite{ostbc}, the decoder can decode the elements in each tuple separately with linear processing. By applying the decoding algorithm to all the tuples and combining them together, the estimated symbols $\tilde{\bm{z}}$ are obtained, which will be further fed to the decoder network to recover the estimated input image. The detailed OSTBC encoding and decoding operations for $N_t = 2, 3$ are given as follows:

The well-known Alamouti scheme \cite{alamouti} is used when the user has two transmit antennas.
To begin with, the latent vector $\bm{z}$ is grouped into pairs: $\bm{z} = \{\{z_1, z_2\},...,\{z_{k-1}, z_k\}\}$, and without loss of generality, the OSTBC design $\mathcal{G}$ for the elements $z_1, z_2$ of the first pair $\bm{z}^{(1)}$ is given as:
\begin{equation}
    \mathcal{G}(\bm{z}^{(1)}) =  \begin{pmatrix} 
    z_1 & -z_2^* \\  
    z_2 & z_1^* 
\end{pmatrix},
\label{eq:alamouti_design}
\end{equation}
and accordingly, we arrange the first antenna to transmit $\sqrt{P}z_1$ and $ -\sqrt{P}z_2^*$ in the first and second time slots, while the second antenna to transmit $ \sqrt{P}z_2$ and $\sqrt{P}z_1^*$, respectively. After passing through the MIMO channel, the received signal $\bm{Y}^{(1)}$ is obtained, whose first and second columns are denoted by $\bm{y}_1$ and $\bm{y}_2$, respectively. According to \cite{alamouti}, we can decouple $z_1, z_2$ from \eqref{eq:received_y} as:
\begin{align}
    \hat{m}_1 &= \sqrt{P}\bm{h}_1^\dagger \bm{y}_1 + \sqrt{P}\bm{y}_2^\dagger \bm{h}_2, \notag \\ 
    \hat{m}_2 &= \sqrt{P}\bm{h}_2^\dagger \bm{y}_1 - \sqrt{P}\bm{y}_2^\dagger \bm{h}_1,
    \label{eq:decouple}
\end{align}
where $\bm{h}_1, \bm{h}_2$ are the first and second columns of $\bm{H}$, and $\hat{m}_1, \hat{m}_2$ are intermediate variables. By substituting the expression of $\bm{y}_1, \bm{y}_2$ into \eqref{eq:decouple}, and applying MMSE estimator to $\hat{m}_1$ and $\hat{m}_2$, the estimated $\hat{z}_1, \hat{z}_2$ are obtained as:
\begin{align}
    \hat{z}_1 &= \hat{m}_1/\left[P(||\bm{h}_1||_2^2+||\bm{h}_2||_2^2)+ \sigma^2 \right],\notag \\ 
    \hat{z}_2 &= \hat{m}_2/\left[P(||\bm{h}_1||_2^2+||\bm{h}_2||_2^2)+ \sigma^2 \right].
    \label{eq:ala_est}
\end{align}
Similarly, the remaining pairs can be estimated forming a sequence $\bm{\hat{z}}$, which will be fed to the subsequent source decoder $g$ to reconstruct the estimate $\widetilde{\bm{X}}$.

We then consider the system with three transmit antennas. Note that when $N_t > 2$, the rate-1 space-time code that achieves full diversity does not exist, and we will adopt two OSTBC designs with rates $1/2$ and $3/4$, respectively. The rate-$1/2$ OSTBC design is given below:
\begin{equation}
\mathcal{G} = \begin{pmatrix} 
z_1 & -z_2 & -z_3 & -z_4 & z^*_1 & -z^*_2 & -z^*_3 & -z^*_4 \\  
z_2 & z_1 & z_4 & -z_3 & z^*_2 & z^*_1 & z^*_4 & -z^*_3 \\ 
z_3 & -z_4 & z_1 & z_2 & z^*_3 & -z^*_4 & z^*_1 & z^*_2
\end{pmatrix}.
\label{eq:rate_half}
\end{equation}
and we partition the latent vector $\bm{z}$ to multiple tuples each consisting of four elements, and let $\bm{z}^{(1)} = [z_1, z_2, z_3, z_4]^\top$ denote the first tuple of $\bm{z}$. After a simple derivation, the input-output relation given in \eqref{eq:received_y} is identical to $\bm{y}_{eff} = \sqrt{P} \bm{H}_{eff}^\dagger \bm{z}^{(1)} + \bm{n}_{eff}$, where $\bm{H}_{eff}$ is given as:
\begin{align}
    \begin{pmatrix} 
\bm{h}_1^\dagger & \bm{h}_2^\dagger & \bm{h}_3^\dagger & \bm{0}^\dagger & \bm{h}_1^\top & \bm{h}_2^\top  & \bm{h}_3^\top & \bm{0}^\top\\  
\bm{h}_2^\dagger & -\bm{h}_1^\dagger & \bm{0}^\dagger & \bm{h}_3^\dagger  & \bm{h}_2^\top & -\bm{h}_1^\top & \bm{0}^\top  & \bm{h}_3^\top \\ 
\bm{h}_3^\dagger & \bm{0}^\dagger & -\bm{h}_1^\dagger  & -\bm{h}_2^\dagger  & \bm{h}_3^\top & \bm{0}^\top & -\bm{h}_1^\top   & -\bm{h}_2^\top \\ 
\bm{0}^\dagger & -\bm{h}_3^\dagger & \bm{h}_2^\dagger &  -\bm{h}_1^\dagger  & \bm{0}^\top & -\bm{h}_3^\top & \bm{h}_2^\top   & -\bm{h}_1^\top \\ 
\end{pmatrix}.
\end{align}
Note that $\bm{h}_j \in \mathcal{C}^{N_r \times 1}$ denotes the CSI vector corresponding to the $j$-th transmit antenna ($j\in \{1,2,3\}$), $\bm{y}_{eff} = [\bm{y}_1^\top, \bm{y}_2^\top, \bm{y}_3^\top, \bm{y}_4^\top, \bm{y}_5^\dagger, \bm{y}_6^\dagger, \bm{y}_7^\dagger, \bm{y}_8^\dagger]^\top$, where $\bm{y}_i$ denotes the $i$-th column of $\bm{Y}^{(1)}$ defined in \eqref{eq:received_y} and $\bm{n}_{eff}$ is obtained in the same way as $\bm{y}_{eff}$.

By left multiplying $\bm{H}_{eff}$ with $\bm{y}_{eff}$, the elements $z_i$ ($i=[1, 4]$) of $\bm{z}^{(1)}$ can be separately decoded as in the Alamouti scheme \eqref{eq:ala_est}, yielding
\begin{align}
    \hat{m}_i &= 2\sqrt{P}\sum_{j=1}^3 {||\bm{h}_j||^2_2} z_i + (\bm{H}_{eff}\bm{n}_{eff})_i, \\
    \hat{z}_i &= \hat{m}_i/\left[ 2P\sum_{j=1}^3 {||\bm{h}_j||^2_2} + \sigma^2 \right],
\end{align}
where $(\cdot)_i$ denotes the $i$-th element of the vector. Then, the decoded sequences will be fed to $g$ to produce the reconstructed image $\widetilde{X}$. Due to the page limit, we refer the interested readers to \cite{ostbc} for a detailed explanation of the design and the corresponding decoding algorithm for the  rate-$3/4$ OSTBC.

\subsection{Multiplexing Scheme}
The previous scheme exploits the diversity of the MIMO system to increase the reliability; however, the orthogonal designs transmit $m$ symbols using $N_u \ge m$ channel uses causing a rate loss that may hinder the system performance when the channel quality is high. Thus, we design a new strategy, called the multiplexing scheme, where $N_t k$ symbols are transmitted in $k$ channel uses. The STM output $\bm{s}$ is identical to the unit-variance latent vector $\bm{z}$ and the transmitter simply reshapes $\bm{z}$ into a matrix, $\bm{Z} \in \mathcal{C}^{N_t \times k}$, and transmits  $\bm{S} = \sqrt{P}\bm{Z}$.

The receiver estimates $\bm{Z}[i], i\in [1,k]$, from $\bm{Y}[i]$ and $\bm{H}$. In traditional digital systems, the latent vector $\bm{Z}[i]$ is comprised of QAM symbols, and sphere decoding can be used \cite{SD}. We do not constrain the transmitted symbols to a finite constellation; thus, we adopt MMSE equalization to first generate a coarse estimation, which is expressed as:
\begin{equation}
    \bm{Z}_{MMSE}[i] = \bm{H}^\dagger \left(\bm{H}\bm{H}^\dagger+\frac{\sigma^2}{P}\bm{I}_{N_r}\right)^{-1} \bm{Y}[i].
    \label{eq:MMSE_multiplex}
\end{equation}
Motivated by \cite{mimo_nos}, we add a residual connection layer, denoted by $\mathrm{Res}$, to calibrate the estimation error of \eqref{eq:MMSE_multiplex} as shown in Fig.~\ref{fig:fig_system} (b). The $\mathrm{Res}$ layer takes the concatenated real-valued $\bm{Y}[i]$ and $\bm{H}$ as input and the final estimation is given by:
\begin{equation}
    \bm{\hat{Z}}[i] = \bm{Z}_{MMSE}[i] + \mathrm{Res}(\bm{Y}[i], \bm{H}).
    \label{eq:residual_multiplex}
\end{equation}
Note that $\bm{\hat{Z}}$ is vectorized and fed to the subsequent decoder $g$ for reconstruction. Both schemes are trained in an end-to-end manner in order to minimize the MSE between the input and output images.

\section{Experiments}

\label{sec:experiment}
\subsection{Training and implementation details}
We evaluate the proposed diversity and multiplexing schemes\footnote{To reproduce the results, we make the source code publicly available at \url{https://github.com/aprilbian/ST_JSCC}.} for the transmission of images from the CIFAR-10 dataset, which consists of 50000 training and 10000 test colored images, each with 32 $\times$ 32 resolution. The number of neurons in the $\mathrm{Res}$ block, denoted by $hid$ is set to 128.  The Adam optimizer is used. The learning rate is initialized at $10^{-4}$ and is reduced by a factor of 0.8 if the validation loss does not drop in 4 consecutive epochs.  Early stopping is used, where the training process will terminate when the validation loss does not drop over 12 epochs. The number of maximum training epochs is set to 400. We only study the cases with $N_t = 2, 3$ and $N_r = 1,2,4$. 

\subsection{Results}\label{sec:sec_tradeoff}

\figtradeoff
We will compare the proposed DeepJSCC approaches with a separation-based bound as a benchmark by assuming that the transmitter knows the instantaneous channel capacity\footnote{Note that in practice, the transmitter does not have CSI, thus it should compress the image at a lower rate to avoid outage\cite{layered_broadcast}.} that could be achieved at each channel block, given by:
\begin{equation}
    C(\bm{H}) = \log_2 \det\left(\bm{I}_2 + \frac{P}{\sigma^2} \bm{H}^H\bm{H}\right).
    \label{eq:MIMO_capacity}
\end{equation}
We then evaluate the average PSNR when BPG compression is used for compressing the image at the rate imposed by $C(\bm{H})$, and average these over the dataset.

We first compare the performance of the  two schemes with the benchmark in Fig.~\ref{fig:fig_tradeoff} for $N_t = 2, N_r \in \{1,2,4\}$ antennas, and a bandwidth ratio of $\rho = 1/8$. We highlight that the CA modules introduced in Sec.~\ref{sec:method} are used to adapt to different SNR regimes ranging from $9$ dB to $17$ dB. As can be seen from the figure, for different numbers of receive antennas, the relative performance between the diversity and multiplexing schemes changes. To be specific, when $N_r = 1$, the diversity scheme is much better than the multiplexing scheme even in the high SNR regime. For $N_r = 2$, the multiplexing scheme outperforms the diversity scheme when $SNR > 16$ dB. For $N_r = 4$, the multiplexing scheme is superior at all SNR regimes. These results are aligned with those in \cite{jsccmimogaussian}, where it is shown that the optimal multiplexing gain increases when the system has more number of antennas, or equivalently, a better diversity-multiplexing trade-off. The intuition is that, when the SNR is low, diversity is more beneficial as it increases the reliability of transmission. However, as the SNR increases, since reliability is warranted, it becomes more beneficial to exploit the transmit antennas for multiplexing to reduce the end-to-end distortion.

We also observe that the DeepJSCC schemes outperform the separation-based benchmark at all SNR regimes when $N_r = 1$. For $N_r = 2$, they are outperformed at higher $SNR$ values and the benchmark always maintains a PSNR gain compared to the learned schemes with $N_r = 4$. We note that the MIMO capacity $C(\bm{H})$ shown in \eqref{eq:MIMO_capacity} increases rapidly with $N_r$, yet it is unachievable with a finite block length. On the other hand, it is remarkable that the proposed DeepJSCC schemes can outperform this idealized upper bound on the separation performance in the low SNR and limited receive antenna scenarios. We note that this is aligned with the previous results on DeepJSCC \cite{bourtsoulatze2019deep}, where it has been observed to provide major gains in the low SNR and low bandwidth ratio regimes. When used for multiplexing gain, multiple antennas effectively increase the available channel bandwidth.


\figNt

Finally, we present the comparison of the diversity and the multiplexing schemes with $N_t = 3$ antennas and a single receive antenna ($N_r = 1$) in Fig.~\ref{fig:fig_nt3}. Two OSTBC designs are evaluated together with the multiplexing scheme. All three models are trained and tested under $SNR \in [2, 10]$ dB. When $\rho = 1/8$, even in the low $SNR$ regime, both diversity schemes are outperformed by the multiplexing scheme. This can be explained by the fact that OSTBC rate for $N_t>2$ is too low to convey the underlying image; hence, it is more beneficial to use the available antennas to improve the multiplexing gain. We highlight that given the $k=384$ MIMO channel uses for $\rho = 1/8$, the length $l$ of the latent vector $\bm{z}$ for the multiplexing scheme is 1152; whereas it is $l = 192$ and $288$ for rate-$1/2$ and rate-$3/4$ OSTBC schemes, respectively, which are too small to achieve good reconstruction quality. On the other hand, when the available bandwidth ratio increases to $\rho = 5/24$, both diversity schemes can achieve comparable or better performance than the multiplexing scheme as $l$ grows larger.  We can also observe that the rate-$1/2$ OSTBC tends to perform better compared to rate-$3/4$ at lower SNR regimes which is intuitive as the rate-$1/2$ code provides larger coding gain, and hence, higher reliability. Though OSTBC offers full diversity, the loss of bandwidth limits the overall end-to-end reconstruction performance when $N_r > 1$,  which is not shown here due to page limitations.

\section{Conclusion}
\label{sec:conclusion}

We presented novel DeepJSCC schemes for wireless image transmission over Rayleigh fading MIMO channels when CSI is available only at the receiver. To the best of our knowledge, these are the first practical JSCC schemes for multiple antenna systems when CSI is not available at the transmitter. In the diversity scheme, the user adopts OSTBC to benefit from the diversity offered by the multiple antennas. For the multiplexing scheme, the user directly maps the codewords to the transmit antennas and the residual assisted MMSE equalizer at the AP is used to estimate the transmitted codewords. Both schemes use end-to-end training to design the outer DeepJSCC encoder/decoder pairs. Simulation results show that, with two transmit antennas, the diversity scheme outperforms the multiplexing scheme with lower system SNR and a smaller number of receive antennas. We also show that the proposed system can outperform the ideal separation-based scheme, which adopts the BPG algorithm and transmits at the MIMO capacity. Exploring the end-to-end reconstruction performance of other diversity-multiplexing trade-off points and the data-driven selection among multiple OSTBCs will be considered in our future work.
\bibliographystyle{IEEEbib}
\bibliography{refs}

\end{document}